\documentclass[twocolumn,showpacs,superscriptaddress,amsmath,amssymb]{revtex4}
\usepackage{hyperref}
\usepackage{graphicx}
\usepackage{dcolumn}
\usepackage{bm}

\begin{document}


\title{Where does a cohesive granular heap break?}

\author{Lyd\'eric Bocquet}
\email{lbocquet@lpmcn.univ-lyon1.fr} \affiliation{ Laboratoire
physique de la mati\`ere condens\'ee et nanostructures UMR CNRS
5586\\ Universit\'e Lyon I -- 43, Bd du 11 Novembre 1918 -- 69622
Villeurbanne Cedex -- France }
\author{Fr\'ed\'eric Restagno}
\email{restagno@lps.u-psud.fr} \affiliation{Laboratoire de
physique des solides -- UMR CNRS 8502 -- Bat. 510 -- Campus
universitaire,\\91405 Orsay Cedex, France}
\author{\'Elisabeth Charlaix}
\email{charlaix@lpmcn.univ-lyon1.fr} \affiliation{ Laboratoire
physique de la mati\`ere condens\'ee et nanostructures UMR CNRS
5586\\ Universit\'e Lyon I -- 43, Bd du 11 Novembre 1918 -- 69622
Villeurbanne Cedex -- France }

\date{\today}

\begin{abstract}
In this paper, we consider the effect of cohesion on the stability of a granular heap
and compute the maximum angle of stability of the heap as a
function of the cohesion. We show that the stability is strongly
affected by the dependence of the cohesion on the local pressure.
In particular,
this dependence is found to determine the localization of the
failure plane. While for a constant cohesive force, slip occurs
deep inside the heap, surface failure is obtained for a linear
variation of the cohesion on the normal stress. Such a transition
allows to interpret some recent experimental results on cohesive
materials.
\end{abstract}

\maketitle

\section{introduction}
The economic impact of particle processing is enormous. Better
methods for the design and synthesis of unit operations involving
divided solids have been identified as a critical need, especially
for pharmaceutical, agrochemicals and specialty chemicals. One of
the important features in industrial processes is to better
understand the transition between a static granular medium and the
avalanche process. Avalanches in non-cohesive granular media have
been extensively investigated \cite{Jaeger88,Rajchenbach90}. A
common characteristic of these studies is that granular motion
occurs in a relative thin boundary layer (around ten grains) at
the surface \cite{Jaeger92} independent of the size of the sample.
On the other hand, recent experiments have explored the relatively
new subject of cohesive granular media such as``humid granular''
media. The presence of capillary bridges between the grains
generate adhesive forces which strongly affect the stability of
the samples \cite{Albert97, Albert99,Hornbaker97, Bocquet98a,
Ursini2002,Fraysse99, Valverde2000,Quintanilla2001}. These
capillary bridges either originate in small amounts of added
fluid, and this situations corresponds to ``lightly wet granular
media''or either are created by a condensable vapor in the
atmosphere \cite{Bocquet98a,Ursini2002,Fraysse99}. In the latter
case the vapor is in chemical equilibrium with the interstitial
liquid bridges. These experiments concern the so-called ``moist
granular media''.

As shown by all these experiments, the behavior of humid granular
media strongly departs from the dry materials and many new
features appear. In a series of experiments on ``lightly wet''
granular media, Tegzes {\em et al.} \cite{Tegzes99}, measured the
angle of repose of a granular media (by the draining crater
method) as a function of the liquid content and of the size of the
system. They observed an increase of the angle of repose with the
liquid content. More precisely, they obtained three regimes: i) at
low liquid content, the ``granular regime'', the avalanche occurs
at the surface and the angle of repose does not depend on the size
of the system; ii) at intermediate liquid content, the
``correlated regime'', the avalanche takes place more deeply in
the heap, the angle of repose depends on the size of the system;
iii) at high liquid content, a kind of ``plastic flow'' is
observed. This various regimes are not accounted for by previous
theoretical analysis.

In numerical and analytical modelizations, cohesion is usually
taken into account as a {\it constant} adhesion force. As an
example, Forsyth {\em et al.} \cite{forsyth01} and Valverde {\em
et al.} \cite{Valverde2000} used a constant van der Waals force to
interpret their experimental results, Olivi-Tran {\em et al.}
\cite{Olivi2002} used a constant capillary force or a constant
force due to a solid bridge in particle dynamics (PD) simulations,
Albert {\em et al.} \cite{Albert97} used a constant capillary
force in a geometrical model and Nase {\em et al.} \cite{nase01},
Mikami {\em et al.} \cite{mikami98} used a constant capillary
force in  PD simulations.

In this paper we show that the assumption of a constant cohesion
is not justified in many practical cases. We then show that the
dependence of the cohesion on normal stress, usually omitted in
the literature, is a key point in determining the stability of a
cohesive granular heap. This effect leads to a dependence of the
stability on the system size, as found experimentally. To this
end, we shall use a continuum analysis \cite{Nedderman} to study
the stability of cohesive sandpiles taking into account a non
constant cohesive stress in the pile. The paper is divided in two
parts. We briefly first show that in a many situations, the
adhesion force between the grains depends on the normal stress.
Then, in the second part, we show how this relationship affects
the stability of a heap and the localization of the slip plane
upon failure.

\section{Contact forces between grains}\label{sec_theo}

The adhesion force between two ideal spheres due to solid-solid
interactions has been calculated in the case of an elastic contact
a long time ago \cite{Johnson71,Derjaguin83,Maugis92a}:
\begin{equation}\label{equ:JKR}
F_{S-S}=f\pi \gamma R
\end{equation}
where $R$ is the sphere radius, $\gamma$ is the solid/intersticial
medium surface tension ($\gamma=\gamma_{SG}$ if the spheres are in
a gas atmosphere, and $\gamma=\gamma_{SL}$ if they are immersed in
a liquid), and $f$ is a numerical factor between 1.5 and 2. The
lower value of $f=1.5$ corresponds to the Johnson-Kendall-Roberts
calculation (JKR) valid  when the attractive forces are strong
enough to deform the spheres surfaces. The higher value $f=2$
corresponds to the Derjaguin-Muller-Toporov (DMT) calculation,
which is valid in the opposite limit of weak attraction and rigid
solids.
The main hypothesis in obtaining Eq. \ref{equ:JKR} is that
surfaces are ideal.
An obvious feature of this result is that adhesion does not depend on
the pressure applied on the surfaces before to separate them
\cite{Israel}. However measured values of the adhesion force
generally departs from the predicted one \cite{Restagno2002}.
As we briefly discuss in the next section, the effect of surface roughness
leads to qualitatively different results, involving a dependence
of the adhesion force on the normal load.

\subsection{Pressure depending adhesion between two
grains}\label{sec}

Restagno {\em et al.} \cite{Restagno2002} have recently measured
the adhesion force between two weakly rough surfaces (a sphere and
a plane) using a surface forces apparatus. The ``pull-off'' force
$F_\text{adh}$ to separate two surfaces has been found to depend
on the a loading normal force, $F$,  with a scaling:
$F_\text{adh}\sim F^{1/3}$.
\begin{figure}[htbp]
  \centering
  \includegraphics[width=7.5cm]{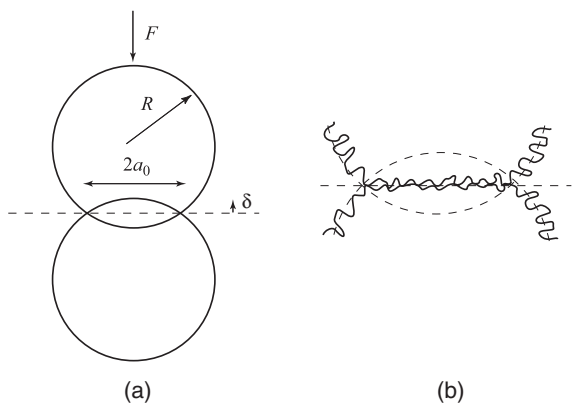}\\
  \caption{(a): Contact between two perfect spheres of radius $R$, Young modulus $E$.
  \\(b): Contact between two rough particules of radius $R$.}\label{fig_hertz}
\end{figure}

The key-point to interpret this result is the roughness of the
surfaces as pointed out 50 years ago by MacFarlane and Tabor
\cite{Bowden50}: the surfaces are in contact only at the tip of
their asperities, and the real contact area $A_\text{r}$ is much
smaller than the apparent contact area $A_\text{app}$ (Fig.
\ref{fig_hertz}. To obtain the adhesion force they used Eq.
\ref{equ:JKR} with an effective surface tension
$\gamma_\text{eff}=\gamma A_\text{r}/A_\text{app}$. On one hand,
different ways can be used to calculate $A_\text{r}$. The first
way is to consider that the pressure in the  molecular contact
area is so high than it can reach the yield stress of the material
\cite{Bowden50}. An other way has been proposed by Greenwood
\cite{Greenwood67} considering the statistical distribution of
heights of the surface asperities. This two models lead to a
simple proportionality between the real contact area $A_r$ and the
normal load:
\begin{equation}\label{equ_tabor}
A_\text{r} \sim F
\end{equation}
On the other hand, the calculation of the apparent contact is the
elastic contact area calculated by Hertz \cite{Landau_elas}~:
$A_\text{app}\sim \left({FR}/{E^*}\right)^{2/3}$, with
$E^*=E/[2(1-\nu^2)]$ the reduced Young modulus of the material,
defined in terms of the Young modulus $E$ and the Poisson modulus
$\nu$ of the material. Gathering results, one obtains eventually
$\gamma_\text{eff} \sim F^{1/3}$, so that Eq. \ref{equ:JKR} is
compatible with the observed experimental scaling.

In the previous experiments, the cohesion between two rough
surfaces due to the direct interaction between the surfaces has
been considered. However, adhesion might also originate in the
presence of a cohesive "binder" in between the surfaces. In many
practical situations, the presence of liquid bridges connecting
the grains leads to strong adhesive forces between the grains
\cite{Albert97,Albert99,Hornbaker97,Bocquet98a,Fraysse99,Ursini2002}.
When the material is placed in an atmosphere of a condensable
vapor \cite{Bocquet98a, Ursini2002}, the size of the liquid bridge
is fixed via chemical equilibrium by the value of external
humidity. The size of such bridges is very small, around a few
nanometers, and one expects that the liquid bridges will condense
only in the regions close to real contact: One expects therefore
that $F_\text{coh} \sim A_\text{r}$, the area of real contact area
(see e.g. \cite{Bocquet98a} for a more complete discussion). Since
as discussed above in Eq. \ref{equ_tabor}, $A_\text{r}$ is
proportional to the external load, one gets eventually:
$F_\text{adh}\sim F$. This picture is however valid when the size
$r$ of the bridges is lower than a typical scale of roughness,
$\ell_R$: In the opposite case, $r>\ell_R$, roughness is not
pertinent any more and the capillary force reduces to a contant
value $2 \pi \gamma R$, with $R$ the radius of the beads, and
$\gamma$ the liquid vapor surface tension.

To sum up, the adhesion force between two surfaces can be written:
\begin{equation}\label{equ_Fn}F_\text{adh}\sim F^n
\end{equation}
with various expected exponents $n$. For humid granular materials,
a transition from from $n=1$ to $n=0$ is expected, depending on
the relative size of the capillary bridges ({\it i.e. } on the
added amount of liquid) compared to roughness~: For small amount
of liquid, the value $n=1$ is expected, while large amounts of
liquid correspond to $n=0$. On the other hand, for dry or liquid
saturated materials in which cohesion is expected to originate in
direct ({\em e.g.} van der Waals) interactions, a value $n=1/3$ is
predicted.

We now derive the local cohesion stress from the adhesion force.

\subsection{from adhesion force to local cohesion stress}

The problem  of the derivation of the interparticle forces $F$
from the bulk stress $\sigma$ is an old problem. Quintanilla {\em
et al.}\cite{Quintanilla2001} have shown, comparing direct
adhesion between particules using an AFM and cohesion stress
measured by determining the tensile stress of a material, that
adhesion increases with the normal load in the material and that a
correlation between the microscopic contact force and a
macroscopic stress measurement can be performed with a good
precision.

\begin{figure}[htbp]
  \centering
  \includegraphics[width=7cm]{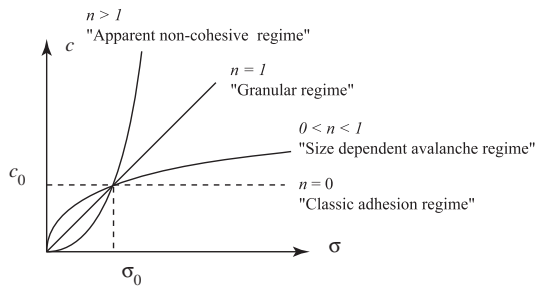}\\
  \caption{different regimes of cohesion in a granular heap.}
 \label{fig_regimes}
\end{figure}

For a system of hard monodisperse spherical particles with a
random isotropic packing, a continuum-like picture leads
to a simple linear relationship between force
$F$ and stress $\sigma$ \cite{NOTE}~:
\begin{equation}\label{equ_cont}
\sigma=\frac{\phi k }{4 \pi R^2}F
\end{equation}
where $R$ is the radius of the particles, $k$ is the coordination
number, which is defined as the average number of contacts per
particle, and $\phi$ is the volume fraction.
Such a relationship has been widely used
in the literature
for powders or wet granular media \cite{Halsey99, Quintanilla2001}.
From
Eq. \ref{equ_cont}, we can deduce the relationship between the
adhesion force $F_\text{adh}$ and the adhesive stress
$c_\text{adh}$ in the granular medium:
\begin{equation}
c_\text{adh} = \frac{\phi k }{4 \pi R^2}F_\text{adh}\sim
\frac{F_\text{adh}}{R^2}
\end{equation}
With the same arguments, we can deduce the relationship between
the normal load and the normal stress:
\begin{equation}
\sigma = \frac{\phi k }{4 \pi R^2}F \sim \frac{F}{R^2}
\end{equation}
Assuming a general relationship of the form of Eq. \ref{equ_Fn},
one gets therefore an adhesive stress which depends on the normal
stress as shown in Fig. \ref{fig_regimes}:
\begin{equation}
c_\text{adh} =c_0\left(\frac{\sigma}{\sigma_0}\right)^n
\label{cohesion}
\end{equation}
where $n$, $\sigma_0$ and $c_0$ characterize the properties of the
material. This relationship is our general starting point to study
the stability of a cohesive heap.


\section{Failure of a cohesive heap}
The most simple approach in the continuum limit to study the
failure of a granular material is the plastic theory of
Mohr-Coulomb. The validity of the continuum approach has been
widely discussed. Nevertheless continuum approaches allow to
predict the Green's function of a granular layer
\cite{reydellet01} and is widely used in soil mechanics
\cite{Nedderman,Halsey99}. We consider the stability of a cohesive
granular heap, characterized by an adhesive stress depending on
the normal stress, as given in Eq. \ref{cohesion} in which we
suppose the Mohr-Coulomb criterion valid. Our aim is to locate in
such a material the "slip plane", where failure occurs. The basis
of our analysis is the Coulomb criterion: a granular material is
stable if {for each surface inside the material} the following
inequality is obeyed:
\begin{equation}
\tau \le \mu \sigma
\label{coulomb}
\end{equation}
where $\tau$ is the shear stress and $\sigma$ is the normal stress. For a cohesive
material, we assume that this condition can be generalized by adding
 in Eq. \ref{coulomb} the adhesive stress
$c_\text{adh}$, as defined in the previous section, to the
normal stress $\sigma$ \cite{Bocquet98a}.

\begin{figure}[htbp]
  \centering
  \includegraphics[width=7cm]{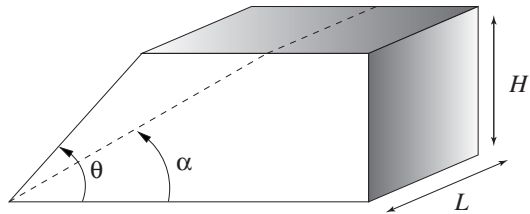}\\
  \caption{Geometry of the granular heap. $\theta$ is the actual slope of the heap,
while $\alpha$ locates the position of a possible failure.}
 \label{fig_heap}
\end{figure}

We consider the geometry depicted in Fig. \ref{fig_heap}: A
granular heap, with height $H$, makes an angle $\theta$ with the
horizontal.  The heap is supposed to be invariant in the direction
perpendicular to the figure. We follow and generalize the approach
given in Ref.  \cite{Nedderman} to the case of a cohesive material
with the law of cohesion given in Eq. \ref{cohesion}. In the
simplified description of Ref. \cite{Nedderman},  the slip surface
is assumed to be planar, making an angle $\alpha$ against the
horizontal.

The force balance in the direction perpendicular and parallel to
the slip plane leads to the following conditions:
\begin{equation}
F=P \sin \alpha \ \ \ N=P \cos \alpha + F_\text{adh}
\end{equation}
where $P$ is the weight of the granular material above the slip
plane; $c=F/S_\text{slip}$ and
$\sigma+c_\text{adh}=N/S_\text{slip}$ are the normal and
tangential stress along the possible slip plane with area
$S_\text{slip}$. In these variables, the stability criterion, Eq.
\ref{coulomb} writes $F<\mu N$. A simple geometric calculation
leads to the relationships \cite{Nedderman}:
\begin{eqnarray}
&P=&{1\over 2} \rho g L H^2 \left( {1\over \tan \alpha}- {1\over \tan \theta}\right) \nonumber \\
&S_\text{slip}=&{H L\over \sin \alpha}
\end{eqnarray}
with $L$ the length of the heap in the invariant direction
(perpendicular to the figure in Fig. \ref{fig_heap}).  Gathering
these results, one gets the following stability criterion:
\begin{equation}
{1\over 2} \rho g  H^2 \left( {1\over \tan \alpha}- {1\over \tan \theta}\right)(\sin \alpha -
\mu \cos \alpha) \le \mu {c_\text{adh} ~H\over \sin \alpha}
\end{equation}
Introducing the angle $\theta_0$ defined as $\tan \theta_0=\mu$,
this relation can be adequately rewritten as:
\begin{equation}
\sin(\theta-\alpha) \sin(\alpha -\theta_0) \le 2\sin \theta_0 \sin \theta {c_\text{adh} \over \rho g H}
\label{criterion1}
\end{equation}

Now, one has to introduce the dependence of the cohesive stress
$c_\text{adh}$ as a function of the normal stress, as discussed in
the previous section. We shall use the general expression given by
Eq. \ref{cohesion}. Using $\sigma=P \cos \alpha/S_\text{slip}$,
one gets after some algebra:
\begin{equation}
c_\text{adh}=c_0 \left( {\rho g H \over 2 \sigma_0}\right)^n \left({\sin(\theta-\alpha)\cos \alpha \over \sin \theta}
\right)^n
\end{equation}
Introducing the cohesion parameter $\zeta_\text{coh}$ defined as:
\begin{equation}
\zeta_\text{coh}={c_0\over \rho g H} \left( {\rho g H \over 2
\sigma_0}\right)^n, \label{def_zeta}
\end{equation}
and the function $f[\alpha]$ defined as:
\begin{equation}
f[\alpha]={(\sin(\theta-\alpha))^{1-n}\over (\cos \alpha)^n} \sin(\alpha -\theta_0)
\end{equation}
Eq. \ref{criterion1} rewrites:
\begin{equation} f[\alpha]\le
\zeta_\text{coh} \sin \theta_0 (\sin \theta)^{n-1}
\label{criterion_cohesion}
\end{equation}
The heap is therefore stable at an angle $\theta$ if this
inequality is verified {\it for all possible $\alpha$} (with $0
\le \alpha \le \theta$). In the opposite case when this inequality
is not verified for some values of $\alpha$, the slip plane
corresponds to the value of $\alpha$ for which this inequality is
``first'' violated. Note that for a non-cohesive material,
$\zeta_\text{coh}=0$, the maximum angle of stability is simply
$\theta_0$.

\subsection{$n\le 1$ case}
First if $\theta\le\theta_0$, the function $f[\alpha]$ is always
negative and the stability condition is trivially verified: For
$\theta<\theta_0$ the heap is always stable.

\begin{figure}[htbp]
  \centering
  \includegraphics[width=6.5cm]{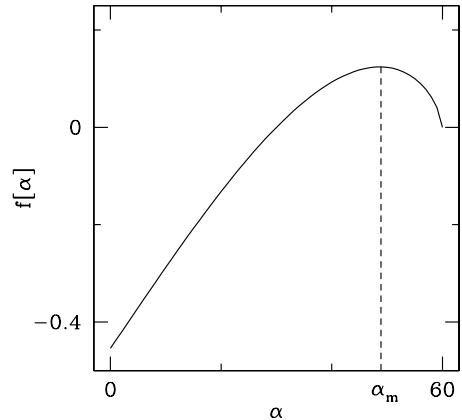}\\
  \caption{Plot of  $f[\alpha]$ as a function of $\alpha$, for a specific choice of parameters:
$\theta=60^o$, $\theta_0=30^o$ and $n=1/3$. The dashed line
locates the position of the maximum $\alpha_\text{m}$.
 }\label{fig_falpha}
\end{figure}

The opposite case $\theta>\theta_0$ is more complex. A typical
plot for this function is then given in figure \ref{fig_falpha}.
This function does exhibit a maximum for a value
$\alpha_\text{m}$, verifying $\partial f/\partial
\alpha(\alpha_\text{m})=0$, and depending of the heap angle
$\theta$. According to criterion \ref{criterion_cohesion}, the
heap is stable if $ f[\alpha_\text{m}]\leq \zeta_\text{coh} \sin
\theta_0 (\sin \theta)^{n-1}$. The maximum angle of stability
$\theta_\text{m}$ thus is solution of the implicit equation:
\begin{equation}
f[\alpha_\text{m}]=\zeta_\text{coh} \sin \theta_0 (\sin
\theta_\text{m})^{n-1}. \label{soltheta}
\end{equation}

A straightforward calculation shows that $\alpha_\text{m}$ obeys
the following relationship:
\begin{equation}
\sin(2\alpha_\text{m} - \theta_0-\theta) \cos \alpha_\text{m}=n
\cos(2\alpha_\text{m} -\theta) \sin(\alpha_\text{m}-\theta_0)
\label{sol}
\end{equation}

It is interesting to consider the two limiting cases $n=1$ and
$n=0$.

\begin{itemize}

\item for $n=1$, the solution of Eq. \ref{sol} is:
$\alpha_\text{m}=\theta$; and $\theta_\text{m}$ is found to obey:
\begin{equation}
{\sin(\theta_\text{m}-\theta_0)\over  \cos \theta_\text{m}
\sin\theta_0}=\zeta_\text{coh}
\end{equation}
This relationship can be rewritten in a more explicit form as~:
\begin{equation}
\tan(\theta_\text{m})=\mu (1+\zeta_\text{coh})=\mu
\left(1+\frac{c_0}{2\sigma_0}\right)
\end{equation}
with $\mu=\tan \theta_0$. This case presents two important
characteristics: (i) failure occurs here at the surface of the
heap, as emphasized by the relationship $\alpha_\text{m}=\theta$;
(ii) whatever the heap height $H$, there is always an effect of
cohesion on the maximum angle of stability. This regime has been
called the {\bf ``granular regime''} according to Tegzes
experiments in which a cohesion effect on the avalanche angle is
found without dependancy on the heap size. In Tegzes experiments,
this regime is found at low liquid content. In section
\ref{sec_theo}, we showed that the value $n=1$ is indeed expected
when cohesion results of small capillary bridges between surfaces
asperities.

\item for $n=0$ (\textbf{``classic adhesion regime''} {\em i.e.}
constant adhesion force), the solution of Eq. \ref{sol} is:
$\alpha_\text{m}=(\theta_0+\theta)/2$, and $\theta_\text{m}$
verifies the equation:
\begin{equation}\label{equ_n0}
{1-\cos(\theta_\text{m}-\theta_0)\over 2 \sin\theta
\sin\theta_0}=\zeta_\text{coh}=\frac{c_0}{\rho g H}
\end{equation}
This is a classical result, as obtained {\em e.g.} in Ref.
\cite{Nedderman}. It is important to note that in this case, (i)
the heap fails deep inside the material, since $\alpha_\text{m}
<\theta_\text{m}$; (ii) the maximum angle of stability
$\theta_\text{m}$ depends on the height of the heap $H$. Such a
dependence is in fact expected when one realizes that for $n=0$, a
"capillary length scale"  can be defined on dimensional grounds~:
$\ell_\text{cap}=c_0/\rho g$ (see eg Eq. \ref{def_zeta}). This
capillary legnth gives the size of a macroscopic element of the
granular medium whose weight equals the adhesion force which act
on it. For instance it maybe the maximum size of a powder aggregat
which can remain stuck under a horizontal surface. Eq.
\ref{equ_n0} shows that the maximum angle of stability of the heap
reduces to $\theta_0$ when the size of the heap is very large
compared to the ``capillary length''.

\item For $n$ in between these two limiting cases, one has to
solve numerically Eqs. \ref{sol} and \ref{soltheta}.
$\alpha_\text{m}$ lies in between $\theta$ and
$(\theta+\theta_0)/2$. We show in Fig. \ref{fig_theta} a typical
result for the dependence of $\theta_\text{m}$ and
$\alpha_\text{m}$ as a function of $\zeta_\text{coh}$ for $n=1/3$.
\begin{figure}[htbp]
  \centering
  \includegraphics[width=7cm]{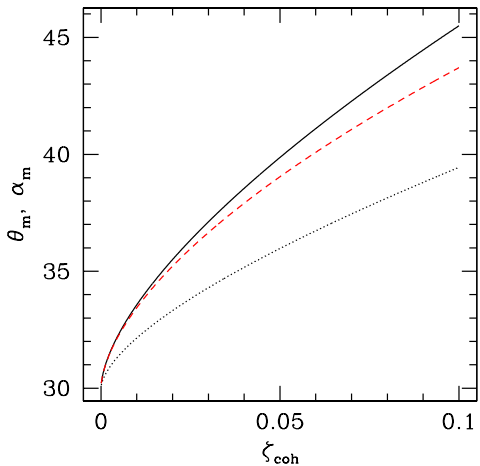}\\
  \caption{Dependence of the maximum angle of stability, $\theta_\text{m}$ (solid line) and failure
plane location, $\alpha_\text{m}$ (dotted line), as a function of
the cohesion $\zeta_\text{coh}$ (angles are here given in
degrees). Physical parameters are: $\theta_0=30^o$, $n=1/3$. The
dashed line is the approximate solution, as given by Eq.
\protect{\ref{approx}}.
 }\label{fig_theta}
\end{figure}
The dependence of $\theta_\text{m}$ in the limit of small cohesion
can be computed analytically. For $\zeta_\text{coh}=0$, one has
$\alpha_\text{m}=\theta_\text{m}=\theta_0$ and one may expand the
angles around this values for small  $\zeta_\text{coh}$:
$\alpha_\text{m}=\theta_0+\delta \alpha$ and
$\theta_\text{m}=\theta_0+\delta \theta$. First linearizing Eq.
\ref{sol}), one gets the relationship $\delta \alpha =\delta
\theta/(2-n)$. Introducing this condition into Eq.
\ref{soltheta}), one gets eventually:
\begin{equation}
\theta_\text{m}-\theta_0=\gamma~ \zeta_\text{coh}^{1\over{2-n}}
\label{approx}
\end{equation}
with $\gamma=(2-n)/(1-n)^{(1-n)/(2-n)}\sin\theta_0
(\cos\theta_0)^{n/2-n}$. Inserting the $H$ dependence of
$\zeta_\text{coh}$, as defined in Eq. \ref{def_zeta}, one gets
therefore
\begin{equation}
\theta_\text{m}-\theta_0\propto H^{n-1\over {2-n}} \label{Hdep}
\end{equation}
This shows in particular that the dependence of the maximum angle
on the heap height is a direct measure of the power $n$,
characterizing the normal stress dependence of the cohesion
stress. This regime is called the {\bf ``size dependent
avalanche'' regime}.
\end{itemize}

\subsection{$n>1$ case}

In this case, it is easy to show that {\it whatever the cohesion},
the maximum angle of stability is $\theta_0$. This regime is
called the {\bf ``apparent non-cohesive'' regime}. First as in the
previous case, when $\theta<\theta_0$, the condition of stability,
Eq. \ref{criterion_cohesion}), is trivially obeyed since the
function $f[\alpha]$ is negative. Now, for $\theta>\theta_0$,
$f[\alpha]$ goes to infinity when $\alpha\rightarrow \theta$. The
heap is therefore always unstable, whatever the cohesion. As a
result, $\theta=\theta_0$ is the maximum angle of stability,
independently of the cohesion $\zeta_\text{coh}$

\section{discussion}

In this paper we have studied the effect of an adhesive stress on
the localization of the slip plane in an avalanche process. More
precisely, we have shown that in a lot of practical case, the
adhesive stress in a material depends on the normal stress. We
have calculated, using a Mohr-Coulomb analysis, the internal angle
of slip $\alpha_\text{m}$.

A few conclusions can be drawn from these results:
\begin{itemize}
\item First, the location of the slip plane, here defined through
$\alpha_\text{m}$, does also strongly depend on this functional
dependence: While for a constant cohesion ({\it i.e. independent
of the normal stress}) the heap slips deep inside, only surface
slip is expected when cohesion is linearly related to normal
stress. \item Second the maximum static angle, $\theta_\text{m}$,
does depend on cohesion via the dimensionless parameter,
$\zeta_\text{coh}$, defined as $\zeta_\text{coh}={c_0\over \rho g
H} \left( {\rho g H \over 2 \sigma_0}\right)^n$. An important
remark is that for $n\ne 1$, the maximum static angle depends on
the height of the heap, $H$. On the other hand, for the specific
value $n=1$ - which is expected in some physical situations (see
discussion in section \ref{sec})- this dependence disappears and
$\theta_\text{m}$ is an intrinsic property of the material,
independently of the geometry. \item Eventually, a change of
regime in the cohesion, {\it e.g.} a change from $n=1$ to $n=0$ as
discussed in section \ref{sec}, will not only modify the
dependence of $\theta_\text{m}$ on cohesion, but more
dramatically, it will change the localization of the slip plane:
For example, while for $n=1$ slip occurs at the surface of the the
heap ($\alpha_\text{m}=\theta_\text{m}$), it will fracture in the
interior of the heap for $n=0$ ($\alpha_\text{m} <
\theta_\text{m}$). In other words, {\it any change of slip
behaviour reflects a transition of cohesion regime}.
\end{itemize}

This features provide a framework to explain the results of Tegzes
{\em et al.}. In view of the previous results, it seems natural to
interprete the different regimes observed in the experiment, as an
indication of a change in the cohesion regime. At low liquid
content, the avalanche takes place at the surface of the heap and
the angle increases linearly with the adhesion. With our approach,
this would correspond to the $n=1$ regime, which is indeed
expected at low liquid content (where the capillary bridges do not
fill the full interstitial space between the grains). At higher
liquid content, a change to a regime with $n<1$ is expected. This
might explain the various observed experimental features, such as
the size dependence.

 As a conclusion, we hope that this
work will motivate further experimental investigation on the
stability of cohesive granular materials. The present results
suggest that a careful determination of the stability properties
and of the failure plane localization yields information on the
cohesion forces between grains.  In particular the heap height
dependence of the maximum angle of  stability should provide a
direct measure of the cohesion properties. Another interesting
geometry is the cylindrical bunker ("Janssen's problem"), in which
cohesion effects, as discussed here, should play a particularly
important role.

\acknowledgements{We thank the R\'egion Rh\^one-Alpes for her
financial support (Programme \textsc{Emergence} 02°1892601).}

\end{document}